\documentclass[11pt, twocolumn, twoside]{IEEEtrantuhh}

\def\BibTeX{{\rm B\kern-.05em{\sc i\kern-.025em b}\kern-.08em
    T\kern-.1667em\lower.7ex\hbox{E}\kern-.125emX}}

\usepackage[dvips]{graphicx}
\usepackage{epsfig,color}
\usepackage{multirow,tabularx}
\usepackage[cmex10]{amsmath}
\usepackage{subfigure}

\def\BibTeX{{\rm B\kern-.05em{\sc i\kern-.025em b}\kern-.08em
    T\kern-.1667em\lower.7ex\hbox{E}\kern-.125emX}}

\newcommand{\smallEq}[2]{\small\begin{eqnarray#2}#1\end{eqnarray#2}}

\begin{document}

\title{Performance Comparison of CP-OFDM and OQAM-OFDM Based WiFi Systems}


\author{
	\small
	\begin{tabular}{cc}
Martin Fuhrwerk, Christoph Thein & Lars H\"aring\\
Institute of Communications Technology (IKT) & Department of Communication Systems\\
Leibniz Universit\"at Hannover, Hannover, Germany & Universit\"at Duisburg-Essen, Duisburg, Germany \\
\{martin.fuhrwerk, christoph.thein\}@ikt.uni-hannover.de & haering@nts.uni-duisburg-essen.de\\
 \end{tabular}
}

\maketitle

\begin{abstract}
In this contribution, a direct comparison of the Offset-QAM-OFDM (OQAM-OFDM) and the Cyclic Prefix OFDM (CP-OFDM) scheme is given for an 802.11a based system. Therefore, the chosen algorithms and choices of design are described and evaluated as a whole system in terms of bit and frame error rate (BER/FER) performance as well as spectral efficiency and complexity in the presence of multi-path propagation for different modulation orders. The results show that the OQAM-OFDM scheme exhibits similar BER and FER performance at a 24\% higher spectral efficiency and achievable throughput at the cost of an up to five times increased computational complexity.

\end{abstract}


\section{Introduction}
\PARstart{T}{he} growing demand for higher data transfer rates in combination with the omnipresence of the need for mobility pushes the limits of today's wireless communication systems. Besides the deployment of multiple antenna schemes, the efficient utilization of the existing resources, time and frequency, can contribute in a significant way to satisfy demands of future wireless communication systems. 
Thereby, frequency agility and in-band and out-of-band spectral efficiency of transmissions can be enhanced by an alternative physical layer modulation scheme called Offset Quadrature Amplitude Modulation (OQAM) Orthogonal Frequency Division Multiplexing (OFDM). It improves the coexistence capabilities of a system with neighboring or narrow-band in-band interferer by its reduced out-of-band emissions and deep spectrum notching features, respectively, due to increased pulse shaping abilities \cite{Viholainen2009}. Furthermore, it exhibits a higher spectral efficiency, compared to other classical multi carrier schemes, due to the absence of an additional cyclic extension as shown in \cite{schaich2010}.

In this contribution, we will quantify the differences in performance, achievable throughput and computational complexity for a IEEE 802.11a standard based system, which is referred to as WiFi system throughout the paper.
The application of the OQAM-OFDM scheme to WiFi systems has already been presented in \cite{Bellanger2008}. The author discusses the advantages of the OQAM-OFDM physical layer and provides an estimate of the achievable bit rate and complexity without a direct comparison to CP-OFDM, which we will deliver in our contribution. Therefore, we base our study on the outcome of the EU project PHYDYAS \cite{phydyas2010}. There the OQAM-OFDM physical layer scheme has been studied and compared to CP-OFDM, focusing on a cellular WiMAX system.

The paper is structured as follows. Section \ref{system design} presents the chosen algorithms used in the system comparison, which is discussed in section \ref{sim results} and focuses on bit and frame error rate performance, spectral efficiency and system complexity. Finally, a summary of the results is given in section \ref{conclusion}.

\section{System Design}\label{system design}


In addition to the common CP-OFDM design of the IEEE 802.11a standard \cite{ieee802.11}, which is used as a reference design within this work, we built an alternative OQAM-OFDM system, based on the polyphase network approach presented in \cite{siohan2002}, to evaluate the bit error rate (BER) and frame error rate (FER) performances in multi-path environments. Therefore the frame structure has been modified such that the synchronization and channel estimation schemes for OQAM-OFDM systems, proposed in \cite{fusco2009} and \cite{Du2009}, respectively, can be used, resulting in an extended preamble compared to the reference design. Nevertheless, the main system parameters, such as number of symbols per frame, are kept the same for both systems to enable a fair comparison.

\vspace{2mm}\subsubsection{Time and Frequency Synchronization} \label{sync}

\begin{figure}[htb]
	\centering
	\includegraphics[width=0.48\textwidth]{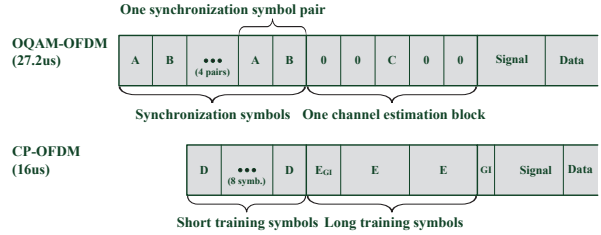}
  \caption{Preamble structures}
  \label{fig:preamble struct} 
\end{figure}

The time and carrier frequency offset (CFO) estimation and synchronization is performed in the time domain using a two step approach. In the first step a coarse timing synchronization, which is based on the modified least-square (MLS) metric presented in \cite{fusco2009} and given in (\ref{eq:tim metr}), is applied for detection of the first symbol start of the frame followed by a coarse CFO correction:
\smallEq{
\label{eq:tim metr}
 \hat{\mu} &=& \underset{\mu}{\mathrm{arg \,max}} \left\{\frac{2|R[\mu]|}{Q[\mu]} \right\} \\
 \Delta \hat{f}[ \hat{\mu}] &=&  \frac{1}{2\pi} \angle\left\{  R[ \hat{\mu}]\right\} 
 }{}
with
\smallEq{
 R[ \hat{\mu}] = \sum_{m}{ y^{*}[m + \hat{\mu}]\, y[m + \Delta m + \hat{\mu}] }
}{}
and
\smallEq{
 Q[ \hat{\mu}] = \sum_{m}{ |y[m + \hat{\mu}]|^2} + |y[m + \Delta m + \hat{\mu}]|^2
}{}
where the correlation width $m$ and the correlation distance $\Delta m$ are defined by
\smallEq{
m \! \in \!\left\{ \!\begin{array}{ll}
\!\left[(\gamma\!-\!2)K\!-\!1,(N_{TR}\!+\!1)K\!-\!1\right] \! & \!\textrm{\footnotesize for OQAM-OFDM}\\
\!\left[0,K-1\right] \! & \! \textrm{\footnotesize for CP-OFDM}
\end{array}\right.
}{*}
and
\smallEq{
\Delta m = \left\{ \begin{array}{ll}
K & \mbox{\footnotesize for OQAM-OFDM} \\
K\left(1+l_\mathrm{CP}\right) & \mbox{\footnotesize for CP-OFDM}.
\end{array}\right.
}{*}
Thereby $\hat{\mu}$ provides the estimated frame start, $\Delta \hat{f}$ indicates the estimated CFO and the number of repeated training symbol pairs $N_\mathrm{TR}$ was chosen to be 6.
$K$, the overlapping factor $\gamma$ as well as $l_\mathrm{CP}$ are specified in Table~\ref{tab:Systemparam}. The selection of $m$ for the OQAM-OFDM system is a trade-off between cross-correlation timing synchronization and auto-correlation CFO estimation performance. In frequency domain the number of allocated carriers for the preamble in the OQAM-OFDM system is 52 and 12 for the CP-OFDM system \cite{ieee802.11}, respectively. 
In the second step the fine frame timing synchronization bases on the cross-correlation scheme utilizing the coarsely time and CFO corrected received signal with the known preamble training sequence. For a CFO greater than 50\% of the subcarrier spacing advanced methods need to be applied to avoid the ambiguity in the phase information which is not considered here. The applied preambles are presented in Fig.~\ref{fig:preamble struct} disregarding the overlapping of OQAM-OFDM symbols and its filter transients in time domain. 

\vspace{2mm}\subsubsection{Channel Estimation and Equalization}

The second section of the OQAM-OFDM preamble, presented in Fig.~\ref{fig:preamble struct}, is used for channel estimation where the symbol, which contains the known information, needs to be guarded by zero-valued symbols to avoid the inherent interference from previous and following symbols. The system design uses a one-tap Zero-Forcing equalizer with the improved channel estimation preamble, as presented in \cite{Du2009}. Other techniques, enabling better performance for channel equalization, e.g. MMSE equalizer as well as multi-tap approaches as in \cite{ihalainen2005} are not considered in this note to compare two basic system designs using similar algorithms. That is why a more compact channel estimation preamble using inherent interference cancelation techniques \cite{Stitz2010} is not introduced.

\vspace{2mm}\subsubsection{Pilot Tones and CFO Tracking}\label{pilots}

Our evaluations have shown that the residual CFO cannot be neglected and needs to be tracked and corrected with the help of pilot tones. The pilot tones, specified in the 802.11a standard and arranged in a comb-like structure, were used for this purpose. The mentioned overlapping of the symbols in time and frequency makes the introduction of auxiliary pilot tones necessary as shown in \cite{Stitz2010}, which mitigates the impact of the system-inherent co-channel and co-symbol interference and enables the use of pilot tones in a similar fashion as in CP-OFDM systems. The pilots are not facilitated for channel tracking due to their wide spacing \cite{troya2003} and the assumption of a static channel for the duration of a frame. 

\section{Results and System Comparison}\label{sim results}

This section holds a performance comparison regarding BER/FER, spectral efficiency, throughput and complexity based on preliminary results derived via Monte-Carlo simulations with $10^3$ channel realizations per SNR value.

\begin{table}[htb]
  \caption{System and simulation parameters}
  \label{tab:Systemparam}
  \scriptsize
  \vspace{-2mm}
  \centering
	\begin{tabular}{p{48mm}||c}
		\hline
		Bandwidth mode																		& 20 MHz \\ \hline
		Sample period $T_S$ 															& 50 ns \\ \hline
		\# of subcarriers $K$ / used carriers $K_u$				& 64 / 52\\ \hline
		\# of data subcarriers / pilots										& 48 / 4 \\ \hline
		Modulation ($M$-QAM)   														& 4 / 16 / 64-QAM \\ \hline
		Code rate $R$ (convolutional code with depth $d$) & \multirow{2}{*}{1/2 (6)}  \\ \hline
		Channel equalization 															& 1-tap Zero-Forcing \\ \hline
		Channel decoding 																	& Hard-bit Viterbi \\ \hline
		Cyclic Prefix overhead $l_\mathrm{CP}$ (CP-OFDM)	& 0.25 \\ \hline
		Prototype filter (OQAM-OFDM)  										& ref. \cite{PHYDYAS2008} \\ \hline
		Overlapping factor (OQAM-OFDM)   									& $\gamma$ = 4 \\ \hline \hline
		Channel model									                		& Hiperlan/2 channel\\
		                                                	& model type A (delay \\
		                                                	& spread of 50ns) \\ \hline
		CFO range (in fraction of the subcarrier spacing)	& \multirow{2}{*}{[-0.1, 0.1]}\\ \hline
		\# of Bytes per frame ($n_\mathrm{bit}$)   		    & 4095 (32760)\\ \hline
	\end{tabular}
	\vspace{-4mm}
\end{table}

\begin{figure*}[htb]
	\centering
		\subfigure{\includegraphics[width=0.499\textwidth]{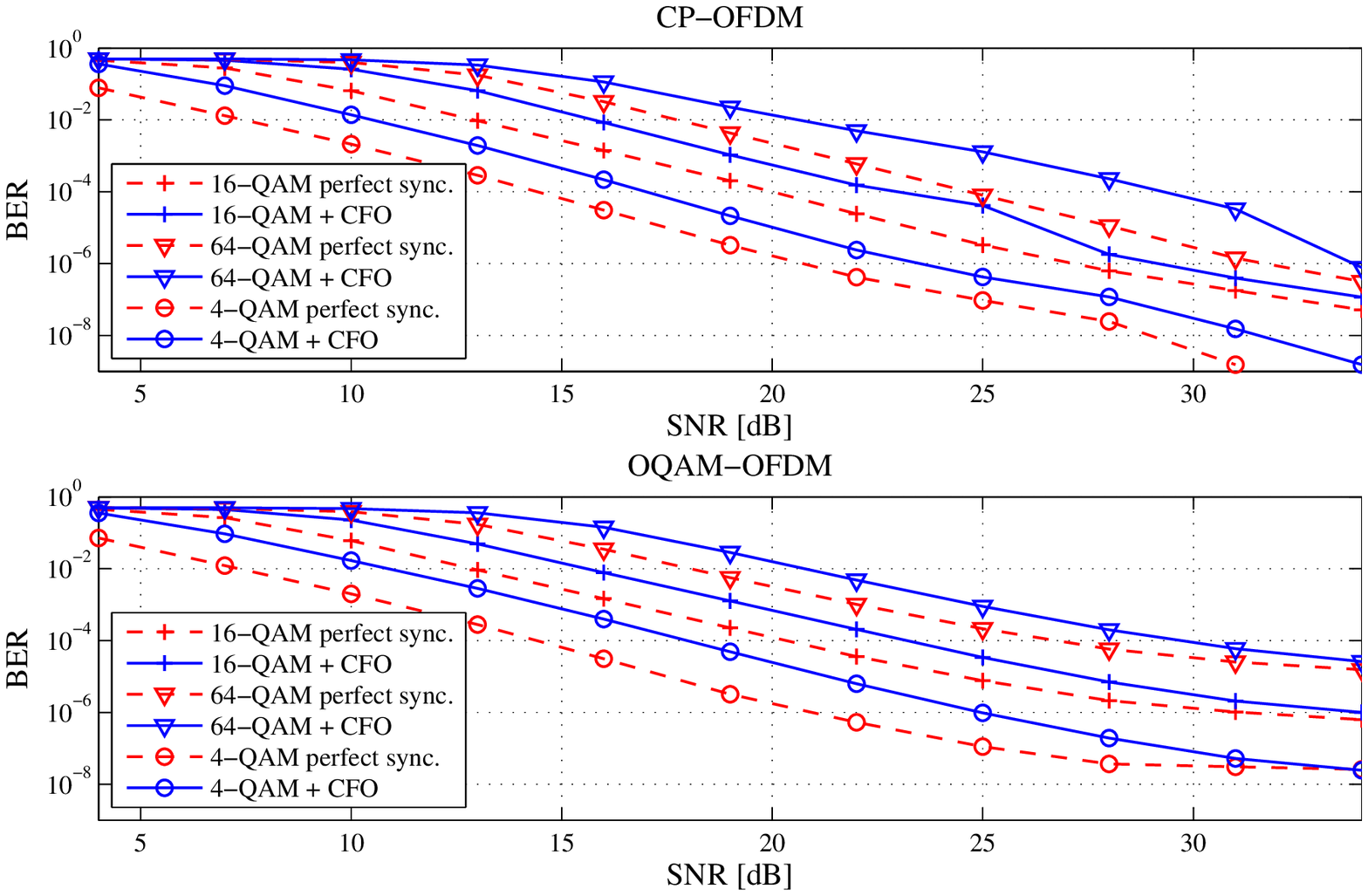}}\hfill
		\subfigure{\includegraphics[width=0.499\textwidth]{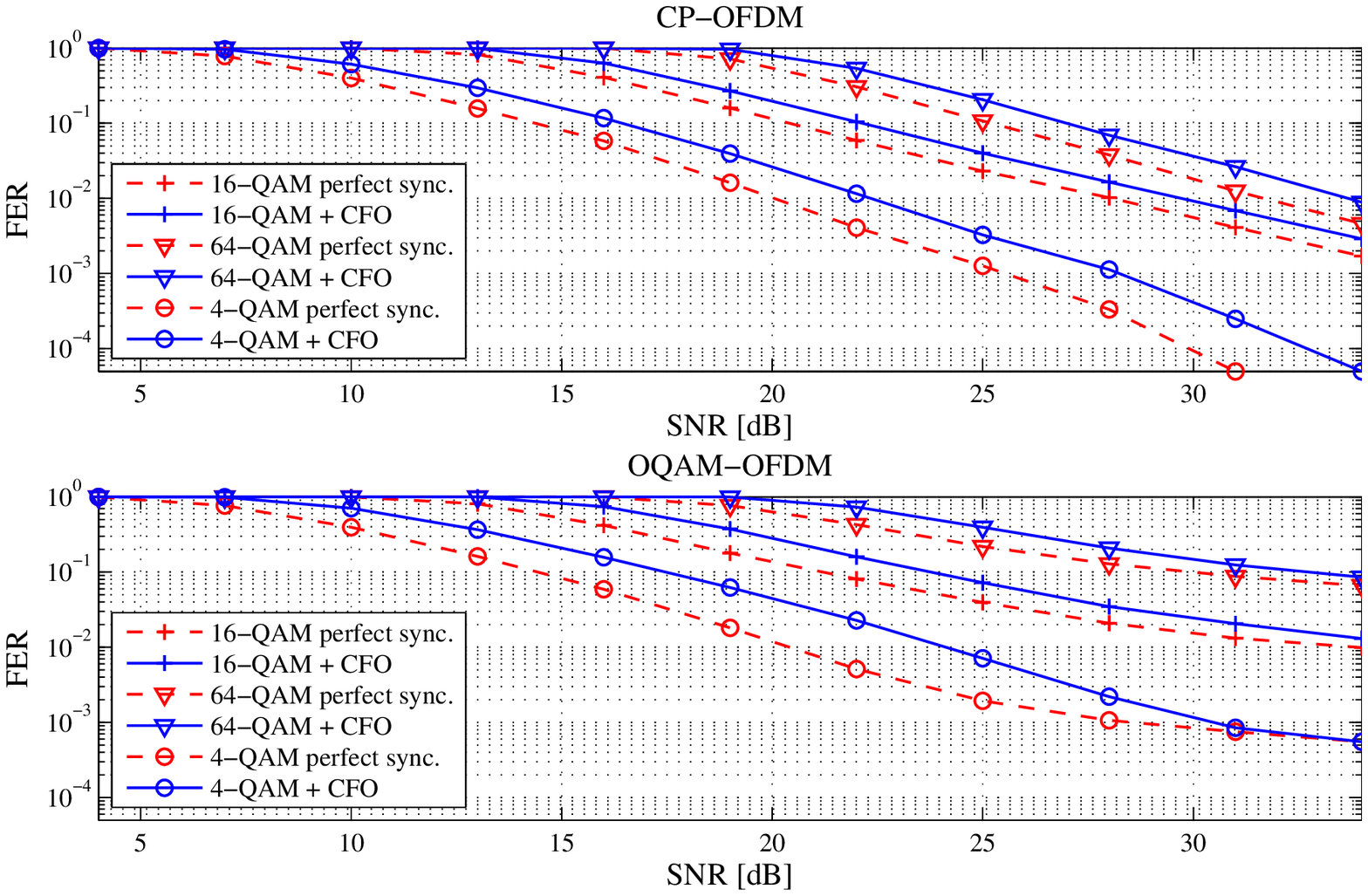}}
	\caption{BER and FER performance in multipath environments (HIPERLAN/2 channel model type A) for different modulation schemes and $R=\frac{1}{2}$}
	\vspace{-2mm}
	\label{fig:sim_BER/FER_all}
\end{figure*}

\subsection{Performance} \label{performance}
\begin{figure}[htb]
	\centering
		\includegraphics[width=0.48\textwidth]{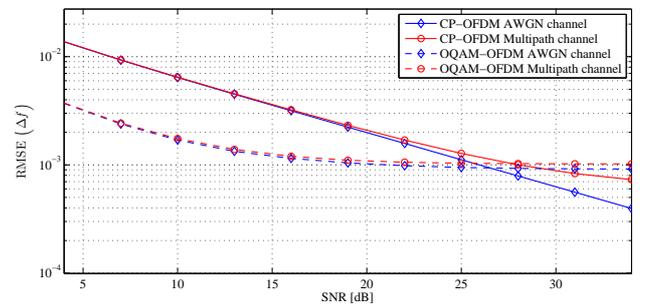}
	\caption{Root Mean Square Error (RMSE) of CFO estimation $\Delta\hat{f}$}
	\vspace{-4mm}
	\label{fig:RMSE_cfo}
\end{figure}

For the presented investigation both systems use the parameters defined in Table~\ref{tab:Systemparam}. For a 4-QAM modulation scheme and compared to the CP-OFDM based system, the OQAM-OFDM scheme shows small degradations of up to 2~dB SNR in the BER performance at the presence of multipath channels, see Fig.~\ref{fig:sim_BER/FER_all}. A significant difference is the BER floor in the OQAM-OFDM system which lies at about $10^{-7}$ for 4-QAM starting at 25~dB SNR. For higher-order modulation schemes (16/64-QAM) the BER floor raises up to about $10^{-6}$ and $10^{-5}$, respectively (Fig.~\ref{fig:sim_BER/FER_all}). A similar behavior can be observed for the FERs with performance floors around $5\cdot10^{-4}$,  $10^{-2}$ and $10^{-1}$ for 4-, 16- and 64-QAM at an SNR of 34~dB resulting in a performance degradation of about 6~dB SNR. These BER and FER floors are induced by intersymbol interference (ISI) due to multipath propagation in combination with the lack of guard intervals.
Another observed detail is the increased robustness of OQAM-OFDM compared to CP-OFDM against CFO in low SNR regions as presented in Fig.~\ref{fig:RMSE_cfo}. This is induced by the applied filter bank together with the fact that the preamble designed for OQAM-OFDM allows a better CFO estimation (refer to section \ref{sync}) during the time domain synchronization. The fast saturating floor for OQAM-OFDM in our system implementation is caused by the selected autocorrelation window size in (\ref{eq:tim metr}).

\begin{table*}[hbt]
	\caption{System comparison for 4095 Bytes payload and code rate $R=\frac{1}{2}$}
	\footnotesize
	\vspace{-2mm}
	\centering
	\begin{tabular}{|c||c|c|c|c|c|}\hline
		\multirow{2}{*}{Type} & \multirow{2}{*}{Modulation Scheme} & Frame length & Throughput & Spectral Efficiency\\ 
		 &  & (ms) & (Mbps) & (bits/s/Hz) \\\hline\hline
		 \multirow{3}{*}{CP-OFDM} & 4-QAM & 2.75 & 11.90 & 0.60 \\\cline{2-5}
		 & 16-QAM & 1.39 & 23.60 & 1.18 \\\cline{2-5}
		 & 64-QAM & 0.93 & 35.15 & 1.76 \\\hline
		 \multirow{3}{*}{OQAM-OFDM} & 4-QAM & 2.24 & 14.66 & 0.73 \\\cline{2-5}
		 & 16-QAM & 1.14 & 28.64 & 1.43 \\\cline{2-5}
		 & 64-QAM & 0.78 & 42.04 & 2.10 \\\hline
	\end{tabular}
	\vspace{-4mm}
	\label{tab:throughput}
\end{table*}

\subsection{Spectral Efficiency and Throughput}

Even though the OQAM-OFDM symbols introduce no cyclic prefix overhead, the overall throughput gain of around 23-24\% is slightly smaller than the bare CP expenditure due to the need of a longer synchronization preamble. The gain in spectral efficiency will decrease further for higher-order modulation and coding schemes and less payload per frame. Table~\ref{tab:throughput} provides a throughput and spectral efficiency $\eta$ comparison using the parameters of Table~\ref{tab:Systemparam} 
 assuming perfect scheduling and transmission conditions. Therefore

\begin{figure}[htb]
	\centering
		\includegraphics[width=0.485\textwidth]{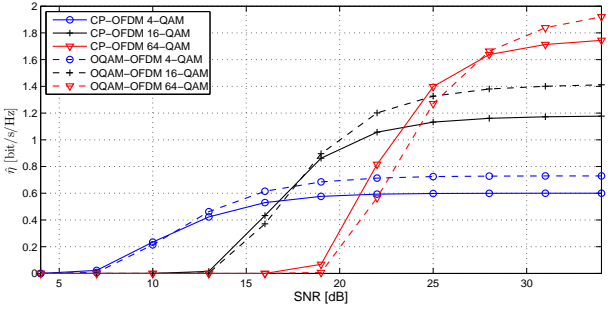}
	\caption{Spectral efficiency $\hat{\eta}$ in presence of CFO and multipath environments (HIPERLAN/2 channel model type A)}
	\vspace{-4mm}
	\label{fig:spectral eff}
\end{figure}

\smallEq{
\label{eq:spec eff}
  \eta = \frac{n_\mathrm{bit}}{B\,\left(T_\mathrm{preamble} + T_S\,{L_\mathrm{data} + T_\mathrm{trans}}\right)}
}{}
with
\smallEq{
\label{eq:n_symb}
\! \! \! \! L_\mathrm{data} \! &=& \! \left\{ \! \begin{array}{ll}
K\,n_\mathrm{symb} \! & \!\mbox{\footnotesize for OQAM-OFDM} \\
K\,n_\mathrm{symb}(1+l_\mathrm{CP}) \! & \! \mbox{\footnotesize for CP-OFDM}
\end{array}\right.\\
\! \! \! \! T_\mathrm{trans} \! &=& \! \left\{ \! \begin{array}{ll}
T_S\,K\left(2(\gamma \!-\!1)\! +\!\frac{1}{2}\right) & \!\textrm{\footnotesize for OQAM-OFDM}\\
0 & \! \textrm{\footnotesize for CP-OFDM}
\end{array} \right.
}{}
and
\smallEq{
n_\mathrm{symb} = \left\lceil 1+\frac{ 16 + 8n_\mathrm{bit} + 6}{48 \, \mathrm{ld}(M) \, R} \right\rceil.
\label{eq:num ofdm sym}
}{}
The preamble duration $T_\mathrm{preamble}$ as defined in Fig.~\ref{fig:preamble struct}, $L_\mathrm{data}$ representing the number of samples needed for the data part of a frame and $T_\mathrm{trans}$ for transmitter filter transient durations and OQAM symbol spread. The calculation of the number of OFDM symbols $n_\mathrm{symb}$ per WiFi frame in (\ref{eq:num ofdm sym}) is derived from \cite{ieee802.11}. In combination with the FERs presented in Fig.~\ref{fig:sim_BER/FER_all} the spectral efficiency for the OQAM-OFDM system does not reach the values presented in Table~\ref{tab:throughput} due to FER floors discussed before. The FER dependend spectral efficiency $\hat{\eta}$ is defined by 
\smallEq{
\hat{\eta} = \eta \left(1-\mathrm{FER}\right)
\label{eq:spec eff FER}
}{}
and presented in Fig.~\ref{fig:spectral eff}.
Referring to \cite{PHYDYAS2008} and \cite{Zhao2011} the suppression of side lobe emission in an OQAM-OFDM system is much better than for CP-OFDM so that the number of needed guard carriers could be relaxed increasing the spectral efficiency even further. A more detailed investigation on this topic is in progress.

\subsection{Complexity}

\begin{figure}[htb]
	\centering
		\includegraphics[width=0.48\textwidth]{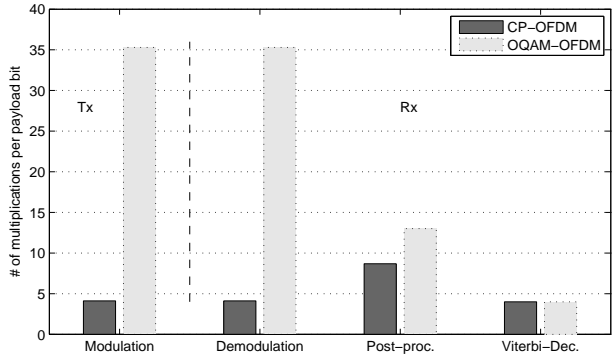}
	\caption{Comparison of number of multiplications per payload bit for $M=4$, $R=\frac{1}{2}$ and 4095 Bytes payload}
	\vspace{-2mm}
	\label{fig:num_mult}
\end{figure}

The main differences between CP-OFDM and OQAM-OFDM are located within the (de-)modulation modules. Considering multiplications as the most costly operations in terms of power and chip area consumption, the following analysis disregards all other operations for simplicity and focuses on the number of multiplications needed to transmit a single data bit. According to \cite{PHYDYAS2009a} and by usage of the Split-Radix algorithm for the calculation of the (I)FFT in combination with (\ref{eq:num ofdm sym}), the modulation complexity $C_\mathrm{mod}$ per transmitted bit, which is equal to the demodulation complexity, is
\smallEq{
\label{eq:eff_mod_oqam}
C_\mathrm{mod}^\mathrm{\{OQAM\}}\! = \!2(2K\! + \!K(\mathrm{ld}(K)\!-\!3)\! + \!4 + \!2\gamma K)\frac{n_\mathrm{symb}}{n_\mathrm{bit}} \! \! \!
}{}
for OQAM-OFDM and
\smallEq{
\label{eq:eff_mod_cp}
C_\mathrm{mod}^\mathrm{\{CP\}} = K(\mathrm{ld}(K)-3)+4)\frac{n_\mathrm{symb}}{n_\mathrm{bit}}
}{}
for CP-OFDM.
Comparing both systems the relative complexity $c_\mathrm{mod}$ can be calculated by 
\smallEq{
\label{eq:eff_ratio_1}
  c_\mathrm{mod} = \frac{C_\mathrm{mod}^\mathrm{\{OQAM\}}}{C_\mathrm{mod}^\mathrm{\{CP\}}}.
}{}
With application of (\ref{eq:eff_mod_oqam}) and (\ref{eq:eff_mod_cp}) $c_\mathrm{mod}$ can be expressed as
\smallEq{
c_\mathrm{mod} &=& \frac{2(2K + K(\mathrm{ld}(K)-3) + 4 + 2\gamma K)}{K(\mathrm{ld}(K)-3) + 4}\\
\label{eq:eff_ratio_2}
  &=& 2 + \frac{4(\gamma + 1)}{\mathrm{ld}(K)-3+\frac{4}{K}}
}{}

so that the additional effort for the OQAM-OFDM based WiFi system would be nearly 8.5 times the complexity of CP-OFDM. In case of a large number of subcarriers the factor decreases significantly. \\
For a fair classification of the performance gain by applying OQAM-OFDM to a WiFi based system the computational complexity of all main system components have to be taken into account, which includes channel (de-)coding, auxiliary pilot calculation, synchronization, channel estimation and equalization as well as phase tracking. The latest three are summarized as post processing in the following. The contribution from encoding, synchronization and auxiliary pilot calculation can be neglected because it is several orders of magnitude smaller than the other examined parts of the systems.
For the $d$-depth hard-bit Viterbi decoder with $n_\mathrm{tail}=d$ tailing bits we consider an implementation which skips multiplications for punctured bits. For this decoder the complexity $C_\mathrm{vit}$ per transmitted data bit is approximated by
\smallEq{
C_\mathrm{vit} &=& \frac{\left\lceil \frac{1}{R}2\left(n_\mathrm{bit} + n_\mathrm{tail}\right)\right\rceil}{n_\mathrm{bit}} \nonumber \\
&\approx & \frac{2}{R} \mbox{, for} \  n_\mathrm{bit} > 20 n_\mathrm{tail}
}{}
assuming the need for two multiplications per received encoded bit.
For the post-processing complexity $C_\mathrm{post}$ we estimate the effort by
\smallEq{
C_\mathrm{post}^\mathrm{\{OQAM\}} &=& \frac{4K_u + 2\cdot4n_\mathrm{symb}\,K_u+4n_\mathrm{symb}\,K_u}{n_\mathrm{bit}} \nonumber \\
&=& 4K_u\left(\frac{3n_\mathrm{symb}+1}{n_\mathrm{bit}}\right)
}{}
and
\smallEq{
C_\mathrm{post}^\mathrm{\{CP\}} &=& \frac{4K_u + 4n_\mathrm{symb}\,K_u+4n_\mathrm{symb}\,K_u}{n_\mathrm{bit}} \nonumber\\
&=& 4K_u\left(\frac{2n_\mathrm{symb}+1}{n_\mathrm{bit}}\right)
}{}
by applying (\ref{eq:num ofdm sym}) and the need of four real multiplications per complex multiplication. Fig.~\ref{fig:num_mult} holds a comparison of both systems itemized to the most expensive parts of a system.
The relative efforts $c_\mathrm{sys}$ for the whole system, $c_\mathrm{sys}^\mathrm{\{Tx\}}$ for transmitter (Tx) side and $c_\mathrm{sys}^\mathrm{\{Rx\}}$ for receiver (Rx) side can be estimated with
\smallEq{
c_\mathrm{sys}^\mathrm{\{Tx\}} &=& c_\mathrm{mod} \\
c_\mathrm{sys}^\mathrm{\{Rx\}} &=& \frac{C_\mathrm{mod}^\mathrm{\{OQAM\}}+C_\mathrm{post}^\mathrm{\{OQAM\}}+C_\mathrm{vit}}{C_\mathrm{mod}^\mathrm{\{CP\}} + C_\mathrm{post}^\mathrm{\{CP\}} + C_\mathrm{vit}}\\
c_\mathrm{sys} &=& \frac{2C_\mathrm{mod}^\mathrm{\{OQAM\}}+C_\mathrm{post}^\mathrm{\{OQAM\}}+C_\mathrm{vit}}{2C_\mathrm{mod}^\mathrm{\{CP\}} + C_\mathrm{post}^\mathrm{\{CP\}} + C_\mathrm{vit}}
}{}
resulting in the values illustrated in Table~\ref{tab:rel_effort}. As it can be seen, the relative system level complexity $c_\mathrm{sys}$ of OQAM-OFDM is reduced to about 3.4 to 4.2 times compared to CP-OFDM instead of about 8.5 times focusing only on the (de-)modulation.

\begin{table}[htb]
	\caption{Relative system complexity for OQAM-OFDM based WiFi System compared to CP-OFDM}
	\centering
	\begin{tabular}{|c||c|c|c|c||c|}\cline{2-6}
	 \multicolumn{1}{l|}{\multirow{2}{*}{}} & \multirow{2}{*}{$M$} & \multirow{2}{*}{$R=\frac{1}{2}$} & \multirow{2}{*}{$R=\frac{2}{3}$} & \multirow{2}{*}{$R=\frac{3}{4}$} & \multirow{2}{*}{$R=1$} \\
	 \multicolumn{1}{l|}{\multirow{2}{*}{}} & & & & & \\ \cline{1-6} \cline{1-6}
	 \multirow{2}{*}{$c_\mathrm{sys}^\mathrm{\{Tx\}}$} & \multirow{2}{*}{all} & \multicolumn{4}{c|}{ \multirow{2}{*}{8.53}} \\
	 & & \multicolumn{4}{c|}{} \\ \hline \hline
   \multirow{3}{*}{$c_\mathrm{sys}^\mathrm{\{Rx\}}$} & 4 & 3.12 & 3.13 & 3.13 & 3.81\\ \cline{2-6}
	   & 16 & 2.73 & 2.75 & 2.75 & 3.87 \\ \cline{2-6}
	   & 64 & 2.47 & 2.49 & 2.50 & 3.93 \\ \hline \hline
   \multirow{3}{*}{$c_\mathrm{sys}$} & 4 & 4.19 & 4.20 & 4.21 & 4.99\\ \cline{2-6}
	   & 16 & 3.71 & 3.73 & 3.74 & 5.06 \\ \cline{2-6}
	   & 64 & 3.36 & 3.39 & 3.40 & 5.13 \\ \hline
	\end{tabular}
	\label{tab:rel_effort}
	\vspace{-4mm}
\end{table}

\section{Conclusion}\label{conclusion}


In this contribution it is shown that the OQAM-OFDM scheme can provide a significant increase in spectral efficiency at the cost of a reasonable higher system complexity which is quantified in terms of number of multiplications. For the modulation schemes 4-QAM and 16-QAM the BER and FER performance is comparable to a CP-OFDM based system leading to a significant increase in spectral efficiency. Due to the shown OQAM-OFDM performance floor the spectral efficiency gain for 64-QAM is extenuated.

\section*{Acknowledgement}
We want to thank Malte Schellmann and Egon Schulz from Huawei Technologies Duesseldorf GmbH, European Research Center for the discussions on complexity and efficiency of wireless communication systems.

\sc
\bibliographystyle{abbrv}

%
%
%

%

\end{document}